\documentclass[aps,prb,reprint,twocolumn,floatfix]{revtex4-1}

\usepackage{amsmath,amssymb}
\allowdisplaybreaks
\usepackage{booktabs}
\usepackage{graphicx}
\usepackage{bm}
\usepackage{here}
\usepackage{subfigure}
\usepackage{newtxtext,newtxmath}
\usepackage{mathtools}
\mathtoolsset{showonlyrefs=true}

\usepackage[dvipsnames]{xcolor}
\graphicspath{{./}{./fig/}}

\begin{document}

\title{Theory of Josephson Current on a Magnetically Doped Topological Insulator}

\author{Tsubasa Toki}
\affiliation{Department of Applied Physics, Nagoya University, Nagoya 464-8603, Japan}
\author{Sho Nakosai}
\affiliation{Department of Applied Physics, Nagoya University, Nagoya 464-8603, Japan}
\author{Yukio Tanaka}
\affiliation{Department of Applied Physics, Nagoya University, Nagoya 464-8603, Japan}
\author{Yuki Kawaguchi}
\affiliation{Department of Applied Physics, Nagoya University, Nagoya 464-8603, Japan}

\begin{abstract}
	Proximity induced superconducting states in the surface of magnetically doped topological insulators can host chiral Majorana modes. We consider a Josephson junction in that system with changing the chemical potential, which drives a topological phase transition in the induced superconducting states as well as a metal-insulator transition in the surface states. The local density of states and the Josephson current are analytically calculated by McMillan's Green's function method in terms of the Andreev reflection coefficient. We show that although the magnitude of the Josephson current is greatly enhanced when the surface state changes from insulating to metallic, its temperature dependence drastically changes at the topological phase transition point, reflecting the appearance of the chiral Majorana modes.
\end{abstract}

\maketitle

\section{Introduction}
It is known that the surface state of a topological insulator (TI) is expressed by the massless Dirac electron with spin-momentum locking~\cite{RevModPhys.82.3045}. This state is protected by time reversal symmetry and can be gapped by magnetization. Also, if we put a superconductor on the surface of a TI, the energy gap opens due to the induced pair potential in this surface state~\cite{PhysRevLett.100.096407}.
It has been predicted that chiral Majorana modes are generated at the boundary between the ferromagnetic insulator (FI) and the superconductor (SC) in FI/SC junctions on the surface of TIs~\cite{PhysRevLett.102.216403}. Then, the tunneling effect and the Josephson effect via chiral Majorana modes in FI/SC junctions have been theoretically studied~\cite{PhysRevLett.102.216404,PhysRevLett.102.107008,PhysRevLett.103.107002,PhysRevB.81.214504,PhysRevB.92.100503}. There are also several theoretical works about charge transport in superconducting junctions on a TI~\cite{PhysRevB.88.075401,PhysRevB.88.121109,PhysicaE.68.107,JETPLett.105.497}. Besides these researches, odd-frequency pairings have been predicted to emerge in such systems~\cite{PhysRevB.87.220506,PhysRevB.92.205424,PhysRevB.96.155426,JPhysCondensMatter.29.295502,PhysRevB.94.134506}.

Up to now, Josephson junctions on the surface of TIs~\cite{NatMater.11.417,PhysRevLett.109.056803,NatCommun.2.575,NatCommun.4.1689,PhysRevB.89.134512,PhysRevLett.114.066801,SupercondSciTechnol.27.104003,NatCommun.8.2019} have been reported, where Bi$_{2}$Se$_{3}$, doped Bi$_{2}$Se$_{3}$, Bi$_{2}$Te$_{3}$, and strained HgTe were used as the TIs. The observed nonsinusoidal current-phase relation of the Josephson current is consistent with the existence of helical edge modes which is a Kramers pair of Majorana modes~\cite{PhysRevLett.114.066801}.
However, it is still difficult to simultaneously stack a ferromagnet and a superconductor on a TI, and FI/SC junctions on the surface of TIs, which supports chiral edge modes, have not been experimentally realized yet.

\begin{figure}[b]
	\centering
	\includegraphics[clip,width=.8\hsize]{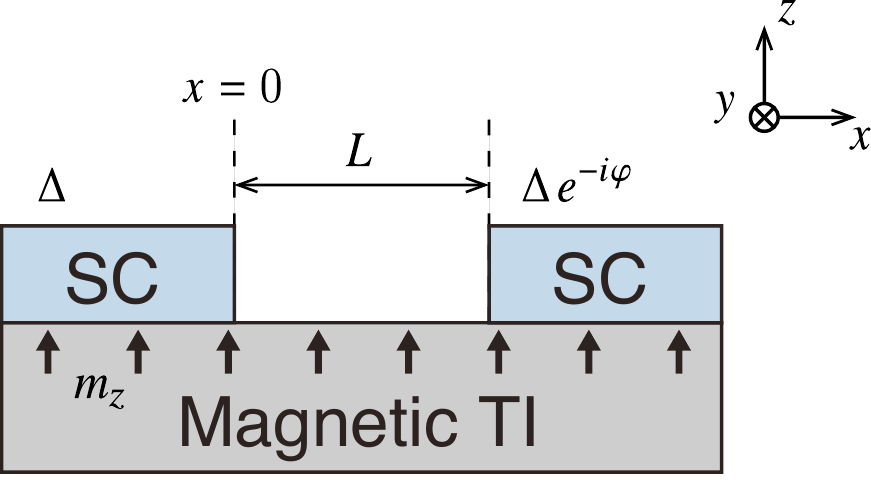}
	\caption{\label{fig:system}Schematic illustration of the system. The surface state of a magnetically doped topological insulator (TI) is coupled to two conventional superconductors (SCs) which are placed with an interval of the length $L$. The system preserves the translational symmetry parallel to the junction (along the $y$ direction).}
\end{figure}

A possible alternative to the FI/SC junction is a superconducting junction on a magnetically doped TI\@. There are several theoretical works on systems with coexisting magnetization and superconducting pairing in the same spatial region~\cite{PhysRevB.92.205424,PhysicaE.87.155,EurPhysJB.90.44,PhysLettA.382.351,PhysRevB.98.024519}. Chiral Majorana modes are predicted to appear in such systems. In this paper, we consider the simplest configuration created just by fabricating superconducting islands on a magnetically doped TI as shown in Fig.~\ref{fig:system}, and discuss how the existence of the chiral Majorana modes affects the Josephson current.
Recently, there is a report on the experimental realization of Josephson junctions on magnetically doped TIs~\cite{CondensMatter.4.1,arXiv180510435}. In magnetically doped TIs, the energy dispersion of the surface Dirac electron has a gap opening when the direction of the magnetization is perpendicular to the surface and its magnitude is larger than the chemical potential of surface Dirac electron $\mu$. In other words, by tuning the chemical potential, metallic (insulating) surface state is realized for $\mu>|m_{z}|$ ($|m_{z}|>\mu$) for $\mu>0$ where $|m_{z}|$ is the magnitude of the out of plane magnetization. At the same time, if we put a superconductor with pair potential $\Delta$ on the magnetically doped TI, topological superconducting state is generated for $|m_{z}| < \sqrt{\mu^{2} + \Delta^{2}}$~\cite{RepProgPhys.75.076501}. Our focus is on the case realized for $\mu < |m_{z}| < \sqrt{\mu^{2} + \Delta^{2}}$ where the chiral Majorana mode is localized at the edge of superconducting region on the TI\@. It is a challenging issue to clarify how the Josephson current flows via chiral Majorana modes in this regime since it is in the insulating phase in the normal state.

\begin{figure*}[t]
	\centering
	\includegraphics[clip,width=.8\hsize]{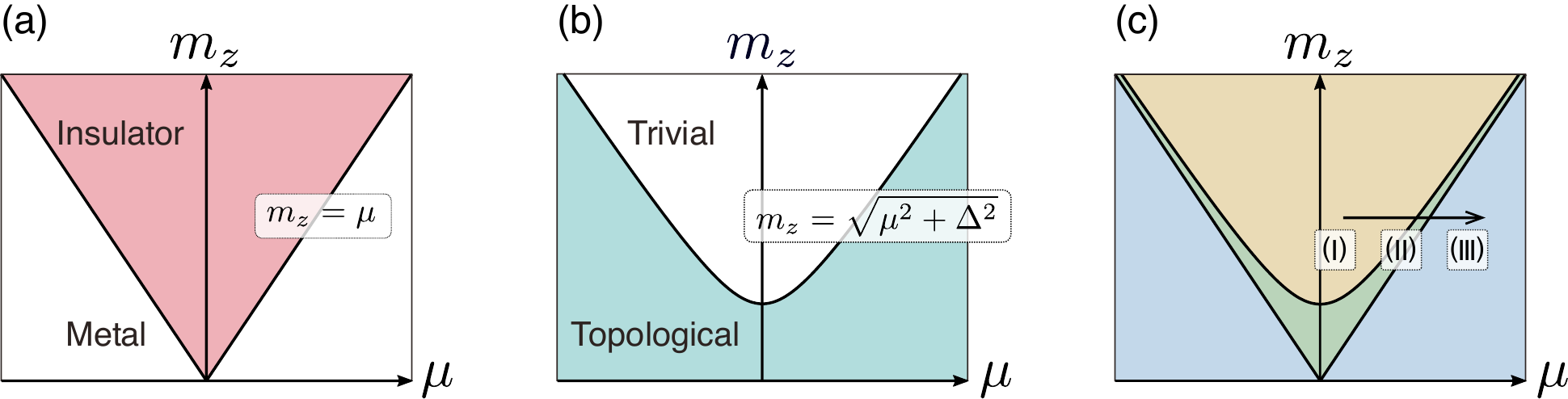}
	\caption{\label{fig:phase_diagram}Phase diagrams of the surface state (a), proximity induced superconducting state (b), and their superposition (c). In (a), the surface of the topological insular in the normal state becomes insulating when magnetic gap is larger than the chemical potential. The induced superconducting state becomes topologically nontrivial provided $|m_{z}| < \sqrt{\Delta^{2}+\mu^{2}}$ as shown in (b). Then, there are three different phases; $\mu<\sqrt{m_{z}^{2}-\Delta^{2}}$, $\sqrt{m_{z}^{2}-\Delta^{2}}<\mu<|m_{z}|$, and $|m_{z}|<\mu$, named Phases (I), (II), and (III), respectively.}
\end{figure*}

Returning to the history of the investigation of the Josephson current~\cite{RevModPhys.76.411}, the Josephson current shows a clear dependence on the presence or absence of zero energy surface Andreev bound states (ZESABSs)~\cite{RepProgPhys.63.1641,SupercondSciTechnol.14.53}. In the junctions with low transmittance at the interface, the temperature dependence of the Josephson current obeys the Ambegaokar-Baratoff relation in the absence of ZESABSs~\cite{PhysRevLett.10.486}. The maximum Josephson current $I_{c}$ saturates with the decrease of temperature $T$, and it is inversely proportional to $R_{N}$, where $R_{N}$ is the resistance in the normal state~\cite{PhysRevLett.10.486}.
On the other hand, in the presence of ZESABSs like the case of $d$-wave superconductor junctions, $I_{c}$ is proportional to $1/T$ in an intermediate temperature regime and proportional to $1/\sqrt{R_{N}}$ at low temperature~\cite{PhysRevB.53.R11957,PhysRevB.56.892,PhysRevLett.77.4070}.
In superconducting junctions on magnetically doped TIs, one can create or annihilate Majorana modes by tuning the chemical potential on the surface of the TI\@. Then, we expect drastic change of the magnitude and the temperature dependence of the Josephson current at the topological phase transition point since Majorana modes are nothing but ZESABSs appearing at the edge of the superconducting regions.

There has been recent development of a theoretical methodology to calculate the Josephson current on a TI based on Green's function of the two-dimensional (2D) Dirac electron system~\cite{PhilosTransRSocA.376.20150246}. This formulation is an extension of McMillan's formulation originally developed for the free electron model with parabolic dispersion~\cite{PhysRev.175.559}.
In this formalism, the Josephson current is expressed by Andreev reflection coefficient like Furusaki-Tsukada formula~\cite{SolidStatCommun.78.299}, which has been developed for $s$-wave superconductor junctions. By an analytic continuation of the Andreev reflection coefficient obtained in the regime of metallic normal state ($|m_{z}|<\mu$), it is possible to calculate the Josephson current even when the normal state is insulating ($|m_{z}|>\mu$). Thus, it is timely to calculate the Josephson current using this formulation in superconducting junctions on magnetically doped TIs.

In this paper, we consider a Josephson junction on the surface of a magnetically doped TI\@. We analytically calculate Green's function on the surface of the TI available for various values of the chemical potential, which drives a topological phase transition in the induced superconducting states and a metal-insulator transition in the surface states. Depending on the relative magnitude of $m_{z}$ to $\mu$, there are three phases: (I) trivial superconductivity and insulating in the normal state for $|m_{z}|>\sqrt{\mu^{2}+\Delta^{2}}$, (II) topological superconductivity and insulating in the normal state for $\sqrt{\mu^{2}+\Delta^{2}} > |m_{z}| > \mu$, and (III) topological superconductivity and metallic in the normal state for $\mu >|m_{z}|$ (see Fig.~\ref{fig:phase_diagram}).
As expected, the magnitude of the Josephson current in Phase (III) is much larger than those in Phases (I) and (II) since the normal state is metallic (insulating) for the former (latter). Besides, we find that the temperature dependence of the Josephson current drastically changes at the phase boundary between Phases (I) and (II), namely, at the topological phase transition point. We focus on the enhancement of the Josephson current at low temperature and evaluate it by the ratio $I_{c}(T=0)/I_{c}(T=0.2T_{c})$, where $I_{\mathrm{c}}$ is the maximum Josephson current as a function of temperature $T$ and $T_{\mathrm{c}}$ is the superconducting transition temperature. The ratio is almost unity in Phase (I), shows rapid increase in Phase (II), and remains large in Phase (III). We attribute the enhancement of the ratio to the chiral Majorana modes appearing in Phases (II) and (III).

The organization of this paper is as follows. In Sec. II, we introduce the model and formulation. In Sec. III, we show the local density of states. In Sec. IV, we show the current-phase relation and temperature dependence of the Josephson current. In Sec. V, we summarize our results.

\section{Model and formulation}
We consider a Josephson junction of two spin-singlet $s$-wave superconductors deposited on the surface of a magnetically doped TI shown in Fig.~\ref{fig:system}. The superconductors provide pair potential in the surface state of the TI via proximity effect while the magnetic moment produces the Zeeman coupling through exchange interaction. Therefore, the Hamiltonian of our interest can read
\begin{align}
	H(\bm x,\partial_{\bm x}) =
	\begin{pmatrix}
		h(\partial_{\bm x}) + M & i \sigma_y \bar{\Delta}(x) \\
		-i \sigma_y \bar{\Delta}(x) & -h^{\ast}(\partial_{\bm x}) - M^{\ast}
	\end{pmatrix}
	\label{eq:hamiltonian}
\end{align}
where $h(\partial_{\bm x}) = v_{\mathrm{F}}(-i\partial_y \sigma_{x} +i\partial_x \sigma_{y}) - \mu$ in the diagonal part is the surface Dirac Hamiltonian with the Fermi velocity $v_{\mathrm{F}}$, the vector differential operator $\partial_{\bm x}=(\partial_x,\partial_y)=(\partial/\partial x,\partial/\partial y)$, and the chemical potential $\mu$. $M = \bm{m}\cdot\bm{\sigma}$ is the Zeeman coupling with $\bm{\sigma}$ being the Pauli matrix vector acting on the spin space. The magnetic moment are finite and fixed in all the region, and we assume it points perpendicular to the surface, i.e., $\bm{m}=(0,0,m_{z})$. Without loss of generality, we assume $m_{z}>0$. The off-diagonal part in Eq.~\eqref{eq:hamiltonian} is the proximity induced pair potential of the spin-singlet $s$-wave form. To consider the junction system depicted in Fig.~\ref{fig:system} we employ $\bar{\Delta}(x)=\Delta (\Theta(-x) + e^{-i\varphi}\Theta(x-L))$ with the interval of the junction $L$ and the phase difference between two superconductors $\varphi$. Although, in general, the induced pair potential depends on the detail of the hybridization and states which the superconductor couples to, we approximate the pair potential has a step-function profile $\Theta(x)$ with a constant magnitude $\Delta$. We choose $\Psi=(c_{\bm{k},\uparrow},c_{\bm{k},\downarrow},c_{-\bm{k},\uparrow}^{\dagger},c_{-\bm{k},\downarrow})^{\mathsf{T}}$ as the basis.

Figure~\ref{fig:phase_diagram} illustrates the phase diagram of the system. The Zeeman term produces a magnetic gap in the surface state. Phases in the normal state or in the region uncovered with the superconductors are shown in Fig.~\ref{fig:phase_diagram}~(a). It becomes metallic when the Fermi energy lies above the magnetic gap and becomes insulating otherwise.
Figure~\ref{fig:phase_diagram}~(b) represents a trivial-topological phase transition in the region covered with the superconductor. According to Ref.~\onlinecite{PhysRevLett.100.096407}, topologically nontrivial superconducting states, which are equivalent to a spinless $ p+ip$ superconductor, are induced on the surface state coupled to a superconductor. When $0 < m_{z} < \sqrt{\Delta^{2}+\mu^{2}}$, the induced pair potential mediates time reversal broken $p$-wave superconductivity~\cite{RepProgPhys.75.076501}. The bulk gap closes along the solid line in the figure, and in the trivial regime, the gap is dominated by the time reversal breaking Zeeman coupling.
We superimpose these two diagrams and obtain Fig.~\ref{fig:phase_diagram}~(c). It consists three phases. Along the arrow in the figure: (I) Josephson junction of topologically trivial superconducting states with an insulating interval; (II) Josephson junction of topological superconductors with an insulating interval; and (III) Josephson junction of topological superconductors with a metallic interval.
In general, topological states host gapless edge modes at their boundary. In the present case, Phases (II) and (III) support Majorana edge modes. Since time reversal symmetry is explicitly broken by the Zeeman coupling, the edge modes are chiral, i.e., a chiral Majorana mode appears at each edge of the two superconductors facing each other across the junction.
However, since the interval of the junction is metallic in Phase (III), the Majorana modes cannot be localized at the edge of the superconductors, and, instead, spread into the interval area. There are also number of propagating modes in the interval which can mediate the current across the junction. In this case, the contribution from Majorana modes to the current becomes less emphasized. The signature of the presence of Majorana modes is strongly manifested in Phase (II). These are explicitly shown in the following sections.

We will follow McMillan's formula~\cite{PhysRev.175.559,RepProgPhys.63.1641} for constructing Green's function to calculate the density of states and the Josephson current. The calculation is performed along Ref.~\onlinecite{PhilosTransRSocA.376.20150246}, where they discuss a similar problem but \textit{without magnetization in the superconducting regions}.
We here briefly summarize the strategy.
In general, Green's function satisfying
\begin{align}
	\left[ E - H(\bm{x}, \partial_{\bm x}) \right] G(\bm{x},\bm{x}^{\prime}) = \delta(\bm{x} - \bm{x}^{\prime})
	\label{eq:hg}
\end{align}
is constructed by the outer product of a solution of the homogeneous differential equation
$\left[ E - H(\bm{x},\partial_{\bm x}) \right] \hat{\psi}(\bm{x})=0$ and that for the conjugate equation $\left[ E - \tilde{H}(\bm{x}^{\prime},\partial_{\bm x^{\prime}}) \right] \hat{\tilde{\psi}}(\bm{x}^{\prime})=0$,
where $\tilde{H}(\bm x,\partial_{\bm x})\equiv H^\mathsf{T}(\bm x,-\partial_{\bm x})$ is the adjoint operator of $H(\bm x,\partial_{\bm x})$ with $\mathsf{T}$ denoting the matrix transpose.
In the present case with the translational symmetry along the $y$ direction,
the Fourier transform of $G(\bm x,\bm x^{\prime})$ with respect to $y$ is given by
\begin{align}
	G(x,x',k_y)=\left\{
		\begin{array}{lll}
			\hat{\psi}_{-}(x)\hat{\tilde{\psi}}^\mathsf{T}_{+}(x^{\prime}) & \textrm{for} & x<x^{\prime}\\
			\hat{\psi}_{+}(x)\hat{\tilde{\psi}}^\mathsf{T}_{-}(x^{\prime}) & \textrm{for} & x>x^{\prime}
		\end{array}\right.,\label{eq:G_general}
\end{align}
where $\hat{\psi}_\pm(x)\ (\hat{\tilde{\psi}}_\pm(x^{\prime}))$ is the eigenstate of $H|_{\partial_y=ik_y}$ ($\tilde{H}|_{\partial_y=-ik_y}$) that satisfies the boundary condition at $x=\pm \infty$.
For the retarded (advanced) Green's function $G^{\mathrm{R (A)}}(x,x',k_y)$, we impose the out-going (in-coming) boundary condition, i.e., we adopt right-going (left-going) electron solution and left-going (right-going) hole solution at $x=+\infty$
and left-going (right-going) electron solution and right-going (left-going) hole solution at $x=-\infty$.
When there are several independent solutions for each of $\hat{\psi}_{\pm}(x)$ and $\hat{\tilde{\psi}}_{\pm}(x)$, the Green's function is written as a linear combination of the outer products of all possible pairs of eigenstates.
The coefficients of the outer products will be determined so as to satisfy the boundary condition at $x=x^{\prime}$.
This boundary condition ensures that the obtained Green's function satisfies Eq.~\eqref{eq:hg}.
For the case of a Josephson junction, by expressing $\hat{\psi}_\pm(x)$ and $\hat{\tilde{\psi}}_\pm(x^{\prime})$
using the Andreev-reflection amplitude, we analytically obtain the Green's function and the resulting local density of states and the Josephson current in terms of these parameters.

Now we move to the concrete calculation for our system given by Eq.~\eqref{eq:hamiltonian}.
To obtain $\hat{\psi}_{\pm}(x)$, we first solve the eigen-solution of $H|_{\partial_{y}=ik_{y}}$.
There are four propagating modes with energy $E$; right-going (in the $x$ direction) electron mode $\hat{A}_{1}e^{ik_{1}x+ik_{y}y}$,
and right-going hole mode $\hat{A}_{2}e^{-ik_{2}x+ik_{y}y}$, left-going electron mode $\hat{A}_{3}e^{-ik_{1}x+ik_{y}y}$, and left-going hole mode $\hat{A}_{4}e^{ik_{2}x+ik_{y}y}$.
The explicit form of the eigenstates of Eq.~\eqref{eq:hamiltonian} in the $x<0$ region are given as
\begin{widetext}
	\begin{align}
		\hat{A}_1 =
		\begin{pmatrix}
			i \\
			\zeta_1 e^{i \theta_1} \\
			- \zeta_1 e^{i \theta_1} \gamma_- \\
			i \gamma_+
		\end{pmatrix},\quad
		\hat{A}_2 =
		\begin{pmatrix}
			i \zeta_2 e^{i \theta_2} \gamma_- \\
			- \gamma_+ \\
			1 \\
			i \zeta_2 e^{i \theta_2}
		\end{pmatrix},\quad
		\hat{A}_3 =
		\begin{pmatrix}
			i e^{i \theta_1} \\
			- \zeta_1 \\
			\zeta_1 \gamma_- \\
			i e^{i \theta_1} \gamma_+
		\end{pmatrix},\quad
		\hat{A}_4 =
		\begin{pmatrix}
			i \zeta_2 \gamma_- \\
			e^{i \theta_2} \gamma_+ \\
			- e^{i \theta_2} \\
			i \zeta_2
		\end{pmatrix},
		\label{eq:eigenstate}
	\end{align}
\end{widetext}
where $\zeta_j$, $\theta_j$ ($j=1,2$), and $\gamma_\pm$ are defined by
\begin{align}
	\zeta_j &= \sqrt{\frac{Z_{j+}}{Z_{j-}}}, \\
	e^{\pm i \theta_j} &= \frac{k_j \pm i k_y}{\sqrt{k_j^{2} + k_y^{2}}},\\
	\gamma_{\pm} &= \frac{\mu\mp m_{z}}{\mu(\mu \pm m_{z})} \frac{\Delta}{E + \Omega_z},
\end{align}
with
\begin{align}
	Z_{j\pm} &= \frac{\mu^{2} - m_{z}^{2} - {(-1)}^{j} \mu \Omega_z + {(-1)}^{j} E m_z}{\mu \pm m_z}, \\
	\Omega_z &= \begin{cases}
		\mathrm{if}\, \mu > m_{z} \\
		\quad \phantom{-} \sqrt{E^{2} - \Delta^{2}_z}   & (E > \Delta_z) \\
		\quad \;i \sqrt{\Delta^{2}_z - E^{2}} & (-\Delta_z \leq E \leq \Delta_z) \\
		\quad -\sqrt{E^{2} - \Delta^{2}_z}  & (E < -\Delta_z) \\
		\mathrm{otherwise} \\
		\quad \phantom{-} \sqrt{E^{2} - \Delta^{2}_z}   & (E > 0) \\
		\quad -\sqrt{E^{2} - \Delta^{2}_z}   & (E < 0) \\
	\end{cases} \\
	\Delta_{z} &= \Delta \sqrt{1 - \frac{m_{z}^{2}}{\mu^{2}}},\\
	k_j &= \frac{\sqrt{Z_{j+} Z_{j-} - (v_\mathrm{F} k_y)^{2}}}{v_\mathrm{F}}.
\end{align}
To obtain normalizable wavefunctions, we need to take $\mathrm{Im}k_{1}\ge0$ and $\mathrm{Im}k_{2}\le0$.
$\Delta_{z}$ is introduced for the short notation, and it also gives an effective gap in the quasiparticle spectrum when $m_{z} < \mu^{2}/\sqrt{\Delta^{2}+\mu^{2}}$. The lowest unoccupied band in the spectrum has double-minima at $k_{\mathrm{min}}=\pm\sqrt{\mu^{2}-m_{z}^{2}(1+\Delta^{2}/\mu^{2})}$ with the gap size $\Delta_{z}$. By increasing $m_{z}$, two minima approach each other, and at $m_{z}=\mu^{2}/\sqrt{\Delta^{2}+\mu^{2}}$ they join up at $k=0$. Then the energy gap becomes $(m_{z}^{2}+\Delta^{2}+\mu^{2}-2m_z\sqrt{\Delta^{2}+\mu^{2}})^{1/2}$, which finally closes at $m_z=\sqrt{\Delta^{2}+\mu^{2}}$ (this is the topological phase transition point in Fig.~\ref{fig:phase_diagram}~(b)).
At $m_z>\mu$, $\Delta_{z}$ becomes pure imaginary, i.e., $\Delta_{z}^2<0$ in the definition of $\Omega_{z}$.

\begin{figure*}[t]
	\includegraphics[width=0.95\hsize]{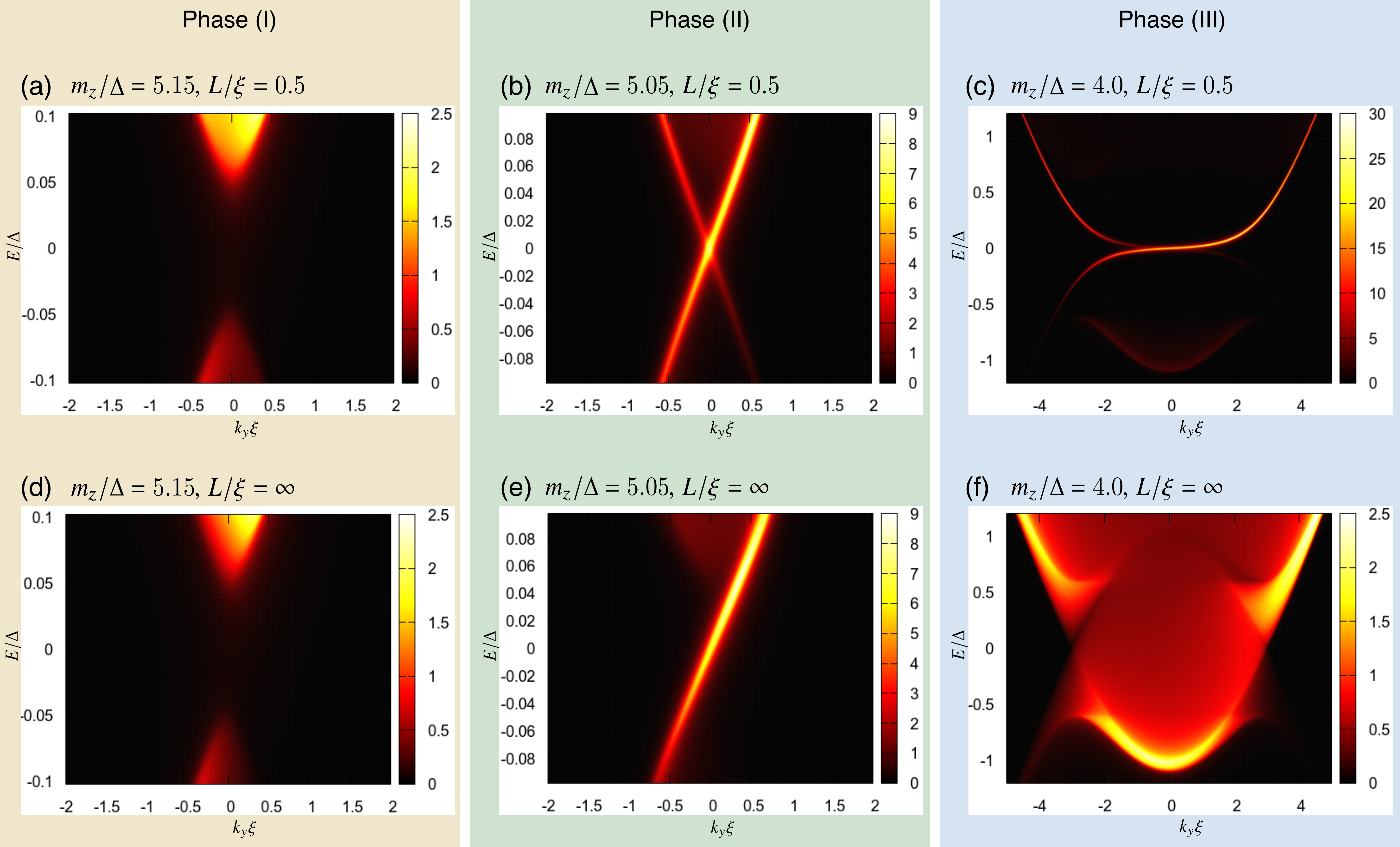}
	\caption{\label{fig:ldos}%
	$k_{y}$ resolved local density of states. Parameters are chosen so that the system is in Phase (I) (a, d), (II) (b, e), and (III) (c, f). The top row is with a narrow interval $L/\xi=0.5$ while the bottom row is with a wider interval $L/\xi=\infty$, i.e., the result for an SN junction. They show gapped surface states in (a) and (d), chiral Majorana mode(s) in (b) and (e), and chiral Majorana modes covered with continuous modes in the interval area of the junction in (c) and (f).
}
\end{figure*}

By solving the scattering problem with the Josephson junction, we obtain two independent solutions that satisfy the out-going boundary condition at $x=+\infty$.
Their analytic forms for $x<0$ is given by
$\hat{\psi}_1(x)=\hat{A}_1e^{ik_1x}+a_1\hat{A}_4e^{ik_2x}+b_1\hat{A}_3e^{-ik_1x}$ and
$\hat{\psi}_2(x)=\hat{A}_2e^{-ik_2x}+a_2\hat{A}_3e^{-ik_1x}+b_2\hat{A}_4e^{ik_2x}$,
where $a_{1(2)}$ and $b_{1(2)}$ are the Andreev-reflection and normal reflection-amplitudes for electron (hole), respectively.
In a similar manner, we calculate two independent solutions for each of $\hat{\psi}_-(x)$, $\hat{\tilde{\psi}}_+(x^\prime)$, and $\hat{\tilde{\psi}}_-(x^\prime)$,
obtaining the retarded Green's function for $x,x^{\prime}<0$ as
\begin{widetext}
	\begin{align}
		G^{\mathrm{R}}(x, x^{\prime}, k_{y}) =
		\begin{cases}
			g_{1} \left[ \hat{A}_{1} \hat{B}_{3}^\mathsf{T} e^{i k_{1} (x-x^{\prime})}
				+ a_{1} \hat{A}_{4} \hat{B}_{3}^\mathsf{T} e^{i k_{2} x - i k_{1} x^{\prime}}
			+ b_{1} \hat{A}_{3} \hat{B}_{3}^\mathsf{T} e^{-i k_{1} (x+x^{\prime})} \right] \\
			\quad + g_{4} \left[ \hat{A}_{2} \hat{B}_{4}^\mathsf{T} e^{-i k_{2} (x-x^{\prime})}
				+ a_{2} \hat{A}_{3} \hat{B}_{4}^\mathsf{T} e^{-i k_{1} x + i k_{2} x^{\prime}}
			+ b_{2} \hat{A}_{4} \hat{B}_{4}^\mathsf{T} e^{i k_{2} (x+x^{\prime})} \right] \quad (x > x^{\prime})
			\\
			g_{1} \left[ \hat{A}_{3} \hat{B}_{1}^\mathsf{T} e^{-i k_{1} (x-x^{\prime})}
				+ \tilde{a}_{1} \hat{A}_{3} \hat{B}_{4}^\mathsf{T} e^{-i k_{1} x + i k_{2} x^{\prime}}
			+ \tilde{b}_{1} \hat{A}_{3} \hat{B}_{3}^\mathsf{T} e^{-i k_{1} (x+x^{\prime})} \right] \\
			\quad + g_{4} \left[ \hat{A}_{4} \hat{B}_{2}^\mathsf{T} e^{i k_{2} (x-x^{\prime})}
				+ \tilde{a}_{2} \hat{A}_{4} \hat{B}_{3}^\mathsf{T} e^{i k_{2} x - i k_{1} x^{\prime}}
			+ \tilde{b}_{2} \hat{A}_{4} \hat{B}_{4}^\mathsf{T} e^{i k_{2} (x+x^{\prime})} \right] \quad (x < x^{\prime})
		\end{cases}.
		\label{eq:green_function}
	\end{align}
\end{widetext}
Here, $\hat{B}_i\,(i=1,2,3,4)$, $\tilde{a}_{1,2}$, and $\tilde{b}_{1,2}$ are the counterparts of
$\hat{A}_i$, $a_{1,2}$, and $b_{1,2}$, respectively, in the conjugate process described by $\tilde{H}|_{\partial_y=-ik_y}$.
More details are given in Appendix~\ref{sect:appendixA}.
In the derivation of Eq.~\eqref{eq:green_function},
we have used the boundary condition at $x = x^{\prime}$:
\begin{align}
	G^{\mathrm{R}}(x,x+0,k_{y}) - G^{\mathrm{R}}(x,x-0,k_{y}) = i v_{\mathrm{F}}^{-1} \begin{pmatrix}\sigma_{y} & 0 \\ 0 & -\sigma_y\end{pmatrix},
	\label{eq:boundary_condition}
\end{align}
obtained by integrating Eq.~\eqref{eq:hg} over $x^{\prime}$ in the section $[x-0,x+0]$,
and $g_1$ and $g_4$ are given by
\begin{align}
	g_{1} &= \frac{i}{2 v_{\mathrm{F}} \zeta_{1} \cos \theta_{1} (1 - \gamma_{+} \gamma_{-})}, \\
	g_{4} &= \frac{i}{2 v_{\mathrm{F}} \zeta_{2} \cos \theta_{1} (1 - \gamma_{+} \gamma_{-})}.
\end{align}

\section{Local density of states}
The local density of states (LDOS) is calculated by
\begin{align}
	&\rho(x,E) \\
	&= \int \frac{d k_{y}}{2\pi} \rho(x,k_{y},E) \\
	&= -\frac{1}{\pi}\int \frac{d k_{y}}{2\pi} \mathrm{Im}\left[ G_{11}^{\mathrm{R}}(x,x,k_{y},E) + G_{22}^{\mathrm{R}}(x,x,k_{y},E) \right].
\end{align}
LDOS at the edge of the left superconductor resolved with $k_{y}$, $\rho(x=0,k_{y},E)$, gives the dispersion of modes. The normal and Andreev coefficients appearing in Green's function in Eq.~\eqref{eq:green_function} are numerically obtained.
They are plotted in Fig.~\ref{fig:ldos} for several sets of $m_{z}$ and the length of the interval $L$. The phase difference between two superconductors and the chemical potential are set to $\varphi = \pi$ and $\mu/\Delta=5.0$, respectively. In Fig.~\ref{fig:ldos}, the panels belong to Phases (I), (II), and (III) from left to right. There are no states observed at zero energy in Fig.~\ref{fig:ldos}~(a) and~(d) whereas there exist modes with linear dispersion at low energy regime in Figs.~\ref{fig:ldos}(b), (c), and~(e). The presence of in-gap states is the consequence of topological property.
From Fig.~\ref{fig:ldos}~(b), it seems two modes propagate to the $+y$ and $-y$ direction. This is because we have two edges of superconductors at $x=0$ and $x=L$, and at each edge one mode is localized. The tail of the wavefunction of the mode at $x=L$ is still large enough at $x=0$ when the length of the interval is comparable with or less than the coherence length $\xi$. This point is explicitly confirmed by looking at the LDOS in an SN junction shown in Figs.~\ref{fig:ldos}~(d-f), which is equivalent to taking $L\to\infty$ in our setting. Now, only the mode propagating to the $+y$ direction remains and the other part disappears (Fig.~\ref{fig:ldos}~(e)) which means we have one chiral Majorana mode localized at $x=0$~\cite{PhysRevLett.100.096407,PhysRevLett.103.107002}. The LDOS of the SN junction also reveals a clear distinction between Phases (II) and (III)\@. When the junction is metallic, there are number of modes at the Fermi energy (see Fig.~\ref{fig:ldos}~(f)). The chiral Majorana modes potentially localized at the edges are no longer distinguishable. It is hybridized with continuous modes in the junction area and spreads into there.

\section{Josephson current}
\begin{figure*}[t]
	\includegraphics[width=0.95\hsize]{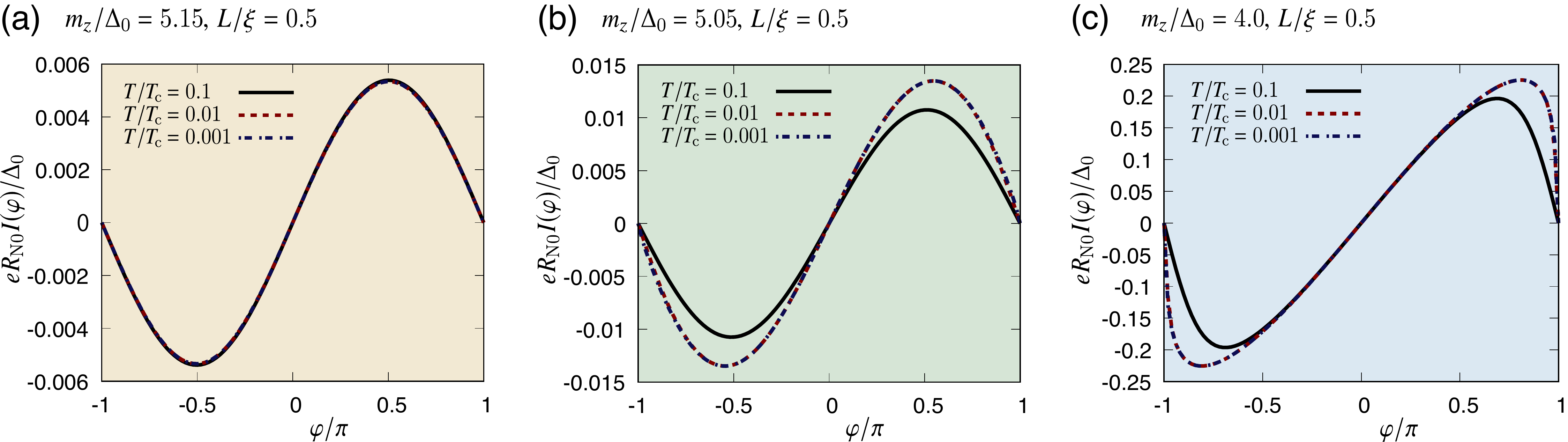}
	\caption{\label{fig:current}%
	Current-phase relation in Phase (I) (a), Phase (II) (b), and Phase (III) (c). They are plotted with varying temperature and the current is measured in units of $\Delta_{0}/e R_{\mathrm{N}0}$ where $R_{\mathrm{N}0}$ is a resistivity in the normal state without magnetization and $\Delta_{0}$ is the pair amplitude at absolute zero. (a) With the insulating junction in the absence of chiral Majorana modes, the relation is in sinusoidal form and has no temperature dependence below $T/T_{\mathrm{c}}<0.1$, being consistent with Ambegaokar-Baratoff behavior. (b) With the insulating junction in the presence of chiral Majorana modes, the relation has similar form to Ambegaokar-Baratoff, though, the critical current keeps growing at low temperature, and it saturates around $T/T_{\mathrm{c}}\sim0.01$. (c) With the metallic junction, the relation and its temperature dependence are the same as Kulik-Omelyanchuk relation.
}
\end{figure*}
We move on to the Josephson current. The analytic continuation $E + i\delta \to i\omega_{n}$ transforms the retarded Green's function to the Matsubara Green's function,  where $\omega_{n}$ is given by $\omega_{n}=2\pi k_{B} T(n + 1/2)$ with an integer $n$. The observables are expressed in the sum over the Matsubara frequency. The Josephson current is derived to be $I = \langle j \rangle + \int^{x}_{0} dx\langle S \rangle$ with the current operator $j	= iev_{\mathrm{F}}	(c_{\downarrow}^{\dagger}c_{\uparrow} - c_{\uparrow}^{\dagger}c_{\downarrow})$ and the source term operator $S = ie\Delta(c_{\downarrow}c_{\uparrow} \allowbreak - c_{\uparrow}^{\dagger}c_{\downarrow}^{\dagger} \allowbreak - c_{\uparrow}c_{\downarrow} \allowbreak + c_{\downarrow}^{\dagger}c_{\uparrow}^{\dagger})$, where $c_{s}^{\dagger}$ ($c_{s}$) is the Fourier transform of the creation (annihilation) operator $c_{\bm{k},s}$ of an electron with spin $s=\uparrow,\downarrow$ with respect to $x$. They can be calculated with respect to the Green's function in the superconductor in $x<0$ as
\begin{align}
	\langle j \rangle &=
	\frac{iev_{\mathrm{F}} k_{\mathrm{B}} T}{2} \sum_{k_{y},i\omega_{n}}
	\left[
		G_{12} - G_{21} + G_{34} - G_{43}
	\right],
	\label{eq:current}
	\\
	\langle S \rangle &=
	ie\Delta k_{\mathrm{B}} T \sum_{k_{y},i\omega_{n}}
	\left[
		G_{14} - G_{23} + G_{32} - G_{41}
	\right].
	\label{eq:source}
\end{align}
After calculation represented in Appendix~\ref{sect:appendixB}, the Josephson current is shown to be
\begin{widetext}
	\begin{align}
		I &= e v_{\mathrm{F}} k_{\mathrm{B}} T \sum_{k_{y},i\omega_n}
		\frac{\Delta (k_{1} + k_{2}) (e^{i\theta_{1}} \zeta_{2} + e^{i\theta_{2}} \zeta_{1})}{4 \mu \sqrt{\omega_{n}^{2} + \Delta_{z}^{2}}}
		\left( \frac{e^{-i\theta_{1}}}{\zeta_{1} \cos\theta_{1}} a_{1} - \frac{e^{-i\theta_{2}}}{\zeta_{2} \cos\theta_{2}} a_{2} \right)
		\\
		&=
		\frac{\Delta (e^{i\theta_{1}} \zeta_{2} + e^{i\theta_{2}} \zeta_{1})}{k_{1} - k_{2}}
		\left( \frac{e^{-i\theta_{1}}}{\zeta_{1} \cos\theta_{1}} a_{1} - \frac{e^{-i\theta_{2}}}{\zeta_{2} \cos\theta_{2}} a_{2} \right),
		\label{eq:josephson}
	\end{align}
\end{widetext}
where we use $k_{1}^{2} - k_{2}^{2} = 4i\mu\sqrt{\omega_{n}^{2}+\Delta_{z}^{2}}$ to get the second equation. In the latter expression, it is obvious that there is no singularity even in the case with $\Delta_{z}$ being a complex number since $\mathrm{Im}k_{1}>0$ and $\mathrm{Im}k_{2}<0$ as we mentioned.

\begin{figure}
	\includegraphics[width=0.8\hsize]{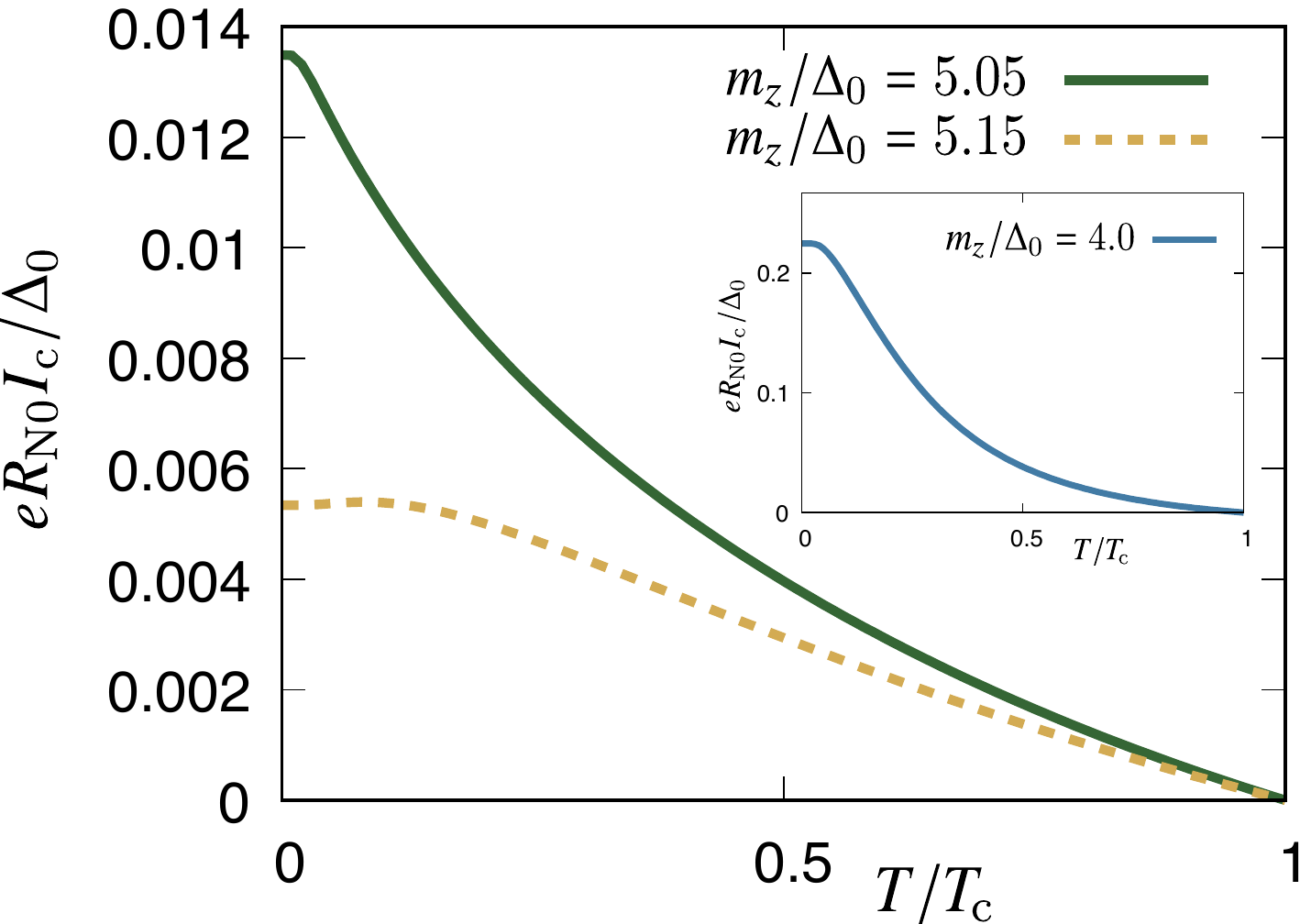}
	\caption{\label{fig:maxcurrent}%
	Maximum value of the Josephson current as a function of temperature. In the main panel, those with an insulating junction are plotted; the yellow dashed line for Phase (I) and the green solid line for Phase (II). The inset show that with a metallic junction, i.e., in Phase (III). We set $L/\xi=0.5$.
	}
\end{figure}

\begin{figure}[t]
	\includegraphics[clip, width=0.8\hsize]{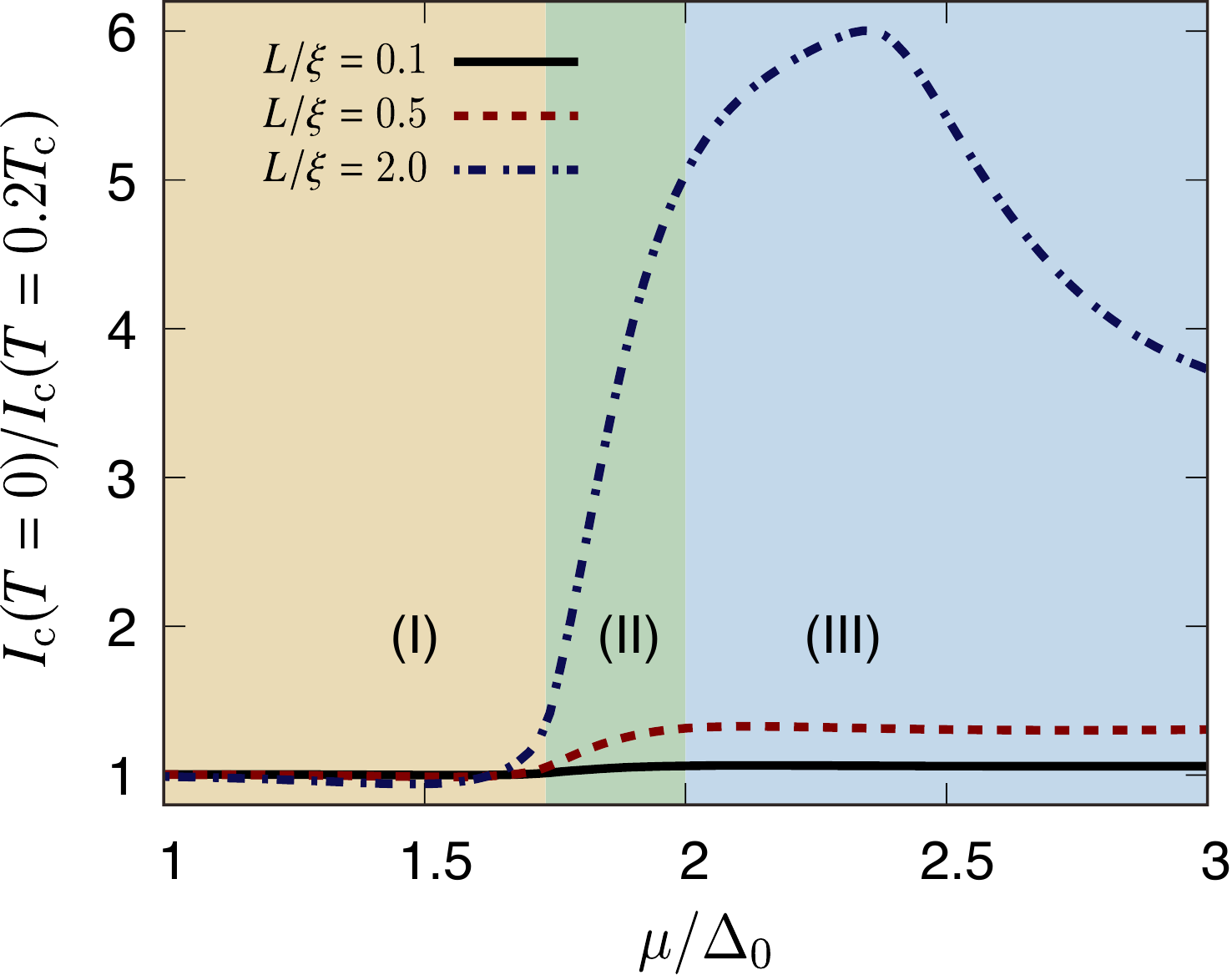}
	\caption{\label{fig:currentRatio}%
	Ratio of the maximum current at zero temperature to that at $T = 0.2T_{\mathrm{c}}$. It is nearly a constant close to unity in Phase (I) and increases in Phases (II) and (III).
}
\end{figure}

\begin{figure}
	\includegraphics[clip, width=0.8\hsize]{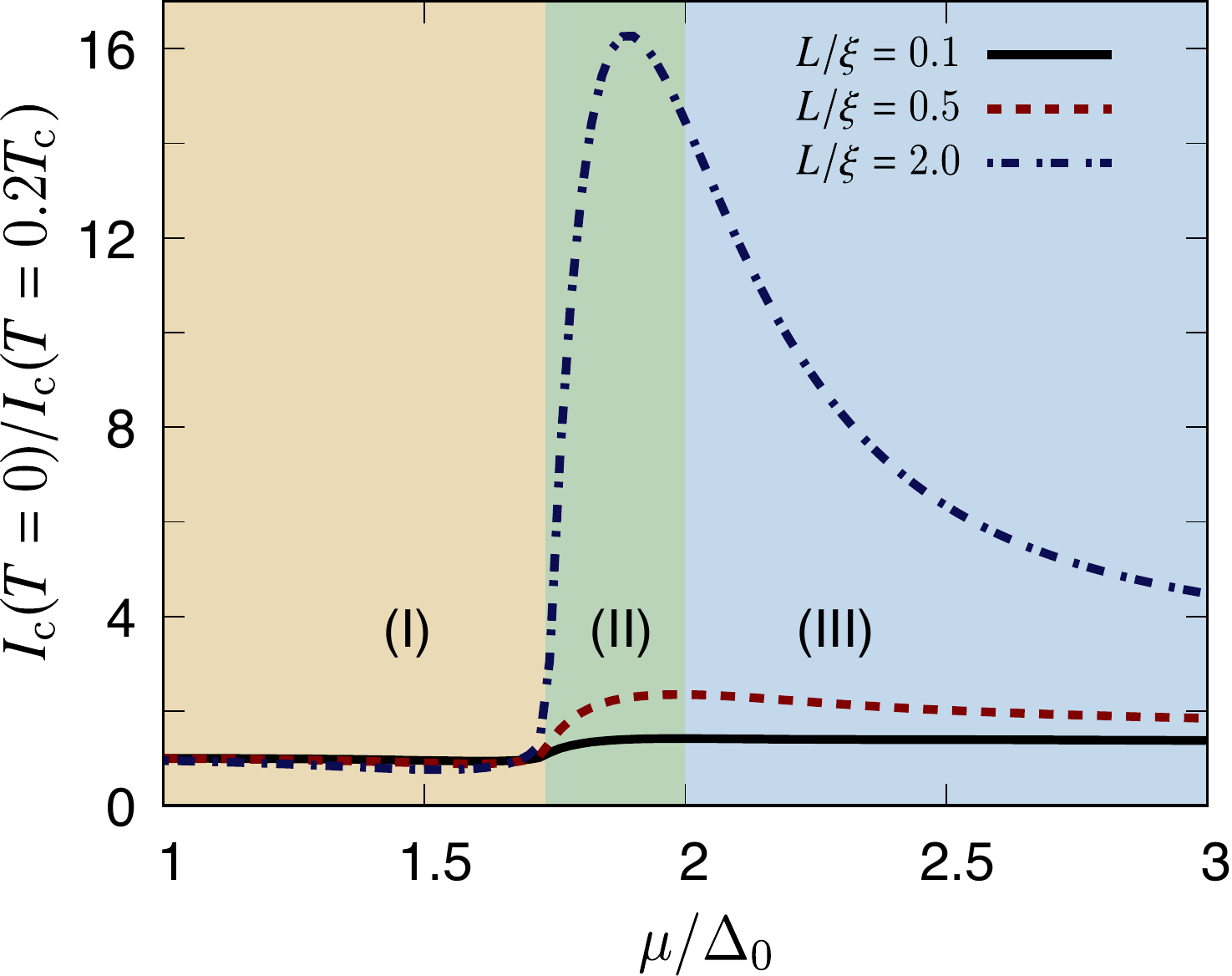}
	\caption{\label{fig:currentRatio1d}%
	Same quantity as in Fig.~\ref{fig:currentRatio} but \textit{in one dimension.}
}
\end{figure}

When the chemical potential is the dominating energy scale, $\mu \gg |E|, \Delta, m_{z}$, so that the quasiclassical approximation is valid, Eq.~\eqref{eq:josephson} can be approximated in a simple form of
\begin{align}
	I = e k_{\mathrm{B}} T \sum_{k_{y},i\omega_{n}} \frac{\Delta_{z}}{\sqrt{\omega_{n}^{2}+\Delta_{z}^{2}}}(a_{1}-a_{2}).
	\label{eq:quasiclassical}
\end{align}
This is an extension of Furusaki-Tsukada formula applied for the Josephson current on magnetically doped topological insulators.

The current-phase relation calculated from Eq.~\eqref{eq:josephson} is shown in Fig.~\ref{fig:current}. The temperature dependence of the pair potential is assumed to be $\Delta(T) = \Delta_{0} \tanh [1.74\sqrt{T_{\mathrm{c}}/T -1}]$. The current is normalized by $\Delta_{0}/e R_{\mathrm{N}0}$ where $R_{\mathrm{N}0}$ is the resistivity of the junction without magnetization in the normal state (We cannot use the resistivity of the junction at $T>T_{c}$ for the normalization, because the normal state is insulating at $m_{z}>\mu$.)
The current-phase relation in Phase (I) has an ordinary sinusoidal form and little temperature dependence below $T/T_{\mathrm{c}}<0.1$ (Fig.~\ref{fig:current}~(a)). This is consistent with Ambegaokar-Baratoff relation, $\sin\varphi \tanh[\Delta(T)/2k_{\mathrm{B}}T]$, obtained in a junction with low transmittance~\cite{PhysRevLett.10.486}.
In Phase (II) (Fig.~\ref{fig:current}~(b)), the behavior of the current-phase relation is nearly in the form of a sinusoidal function. When the temperature goes down, however, the maximum Josephson current grows and its position is slightly away from $\varphi=\pi/2$.
These properties are different from the naively expected Ambegaokar-Baratoff relation. Based on the result of LDOS (Fig.~\ref{fig:ldos}~(b)), these behavior can be attributed to the chiral Majorana modes arising in the superconducting gap.
On the other hand, in Phase (III) (Fig.~\ref{fig:current}~(c)), the maximum current flows when the phase difference is close to $\pi$. The behavior is similar to superconductor-short normal metal-superconductor junctions with fully transparent interface, known as Kulik-Omelyanchuk relation~\cite{FizNizkTemp.3.945}, or unconventional superconductor-short junctions with low transmission in the presence of ZESABSs~\cite{PhysRevB.53.R11957,PhysRevB.56.892}.
Figure~\ref{fig:maxcurrent} shows the temperature dependence of the maximum value of the Josephson current $I_{\mathrm{c}}$. The maximum current saturates near zero temperature in all cases. It should be noted again that the surface state without superconductor attached is insulating in Phases (I) and (II). It is then natural that the saturated value in Phases (III) is much larger than those in Phases (I) and (II). If we closely look at the temperature dependence, however, we notice that the saturation occurs relatively higher temperature when the superconducting state is trivial, whereas the maximum current keeps getting larger till close to zero temperature when the superconducting state becomes topological.

To extract these increase at low temperature, we take the ratio between the current at zero temperature and at a reference temperature, which we choose as $T = 0.2 T_{\mathrm{c}}$. Figure~\ref{fig:currentRatio} shows the ratio as a function of the chemical potential. By increasing the chemical potential, the system goes through two transition points around $\mu/\Delta_{0}=1.75$ and $2.0$. When the surface state is insulating, i.e., with small $\mu$, the current shows saturation below the reference temperature. The ratio significantly increases when the superconducting regions become topological and chiral Majorana modes appear. It keeps increasing until the system is in the metallic phase and then decreases converging to a constant.
To emphasize the contribution of Majorana modes, we also calculate the ratio in the one dimensional space, which is illustrated in Fig.~\ref{fig:currentRatio1d}. This case corresponds to taking only the contribution from $k_{y}=0$ in 2D case. Now the peak locates in Phase (II), which indicates that the increase of the Josephson current is mostly attributed to the presence of Majorana modes not to continuous modes in metallic surface states.
The shift of the peak in 2D space can be understood as follows. As shown in Fig.~\ref{fig:ldos}, the energy of chiral Majorana modes depends on the wave number parallel to the junction, and it is on $E = 0$ only when $k_{y} = 0$. Therefore, at low temperature comparable to the finite energy of chiral Majorana modes, the contribution of Majorana modes to the Josephson current is overwhelmed by the contribution from the continuous modes.

\section{Summary}
In this paper, we have studied charge transport of the Josephson junction on the surface of magnetically doped TIs. We have analytically obtained the Green's function on the surface of TIs available for the various chemical potential, which drives a topological phase transition in the induced superconducting states and a metal-insulator transition in the surface states of 2D Dirac electron. There are three parameter regions: (I) trivial superconductivity with insulating normal state for $|m_{z}|>\sqrt{\mu^{2}+\Delta^{2}}$, (II) topological superconductivity with insulating normal state for $\sqrt{\mu^{2}+\Delta^{2}} > |m_{z}| > \mu$, and (III) topological superconductivity with metallic normal state for $\mu > |m_{z}|$.
It is remarkable that non-zero Josephson current is obtained even for (I) and (II) though it is rather small. By focusing on the ratio $I_{c}(T=0)/I_{c}(T=0.2T_{c})$, we have distinguished three regions clearly. This ratio is almost unity in Phase (I), suddenly increases in Phase (II), and remains enhanced in Phase (III). Figure~\ref{fig:currentRatio} tells that the existence of chiral Majorana states can be detected by an experimental observation of a prominent increase in the Josephson current at low temperature in the insulating phase (Phase (II)).
The current available experimental technique would allow us to tune the chemical potential widely enough to go across these phase boundaries. Thus, the observation of the temperature dependence of the Josephson current provides detection of Majorana modes.

In this paper, we have calculated Josephson current in the ballistic regime. In the actual surface state of TIs, the impurity scattering effect also influences on the charge transport. It is interesting to clarify the possible odd-frequency spin-triplet $s$-wave pairing which is robust against the impurity scattering~\cite{PhysRevLett.98.037003,JPSJ.81.011013}.
In addition to this point, one of the remarkable finding of this paper is the presence of Josephson current in the regime where electronic property of surface state of TIs in the normal state is insulating.
There are several studies on charge transport in superconducting quantum anomalous Hall system where the normal state is insulating. One can see our model as the proximity induced superconducting state in an anomalous quantum Hall state. The magnetic gap in the surface Dirac state of a TI provides a half quantized Chern number $\sigma=1/2$, which have been intensively investigated both in theories~\cite{PhysRevB.82.184516,PhysRevB.83.224524,PhysRevB.86.174512,PhysicaE.55.13,NatCommun.5.3232} and experiments~\cite{Science.357.294}.
Besides the setup shown in Fig.~\ref{fig:system}, our system may be realized at the interface between a 2D TI and an FI with a superconducting junction deposited on it, or same configuration at hinge states in higher order TIs~\cite{Science.357.61}.

\begin{acknowledgments}
	The authors would like to thank R. Yano, S. Kashiwaya, B. Lu, and P. Burset for fruitful discussion. This work was supported by JST CREST Grant No.\@ JPMJCR16F2,
	and JSPS KAKENHI Grant Nos.\@ JP15H05851, JP15H05853, JP15K17726, JP18H01176, and JP19H01824.
\end{acknowledgments}

\appendix
\section{McMillan's Green's function method}\label{sect:appendixA}
Here we give all the detail of the calculation for constructing McMillan's Green's function.
\subsection{Scattering solutions}
Solving the scattering problem with the Josephson junction results in the following four independent wavefunctions in the $x<0$ region~\cite{PhysRevB.92.205424}:
\begin{widetext}
	\begin{align}
		\hat{\psi}_1(x) &= \hat{A}_1 e^{i k_1 x} + a_1 \hat{A}_4 e^{i k_2 x} + b_1 \hat{A}_3 e^{-i k_1 x}, \\
		\hat{\psi}_2(x) &= \hat{A}_2 e^{-i k_2 x} + a_2 \hat{A}_3 e^{-i k_1 x} + b_2 \hat{A}_4 e^{i k_2 x}, \\
		\hat{\psi}_3(x) &= c_3 \hat{A}_3 e^{-i k_1 x} + d_3 \hat{A}_4 e^{i k_2 x}, \\
		\hat{\psi}_4(x) &= c_4 \hat{A}_4 e^{i k_2 x} + d_4 \hat{A}_3 e^{-i k_1 x}.
	\end{align}
    The former (latter) two satisfy the out-going boundary condition at $x=+\infty$ ($x=-\infty$).
    $c_3$ ($d_4$) and $d_3$ ($c_4$) are the transmission coefficients for electron and hole, respectively, when electron (hole) is injected from the right.
	On the other hand, the eigenstates of the conjugate processes obtained by solving
	\begin{align}
		\tilde{H}({\bm x}^{\prime},\partial_{\bm x^{\prime}}) =
		\begin{pmatrix}
			h(\partial_{x'},-\partial_{y'}) + M^\ast & i \sigma_y \Delta(x') \\
			-i \sigma_y \Delta(x') & -h^{\ast}(\partial_{x'},-\partial_{y'}) - M
		\end{pmatrix}
		\label{eq:conjugate_process}
	\end{align}
	are in the form of $\hat{B}_{1} e^{i k_{1} x^{\prime} - i k_{y} y^{\prime}}, \hat{B}_{2} e^{-i k_{2} x^{\prime} - i k_{y} y^{\prime}}, \hat{B}_{3} e^{-i k_{1} x^{\prime} - i k_{y} y^{\prime}}, \hat{B}_{4} e^{i k_{2} x^{\prime} - i k_{y} y^{\prime}}$ with
	\begin{align}
		\hat{B}_{1} =
		\begin{pmatrix}
			i e^{-i \theta_1} \\
			\zeta_{1} \\
			- \zeta_{1} \gamma_{-} \\
			i e^{- i \theta_{1}} \gamma_{+}
		\end{pmatrix};
		\hat{B}_{2} =
		\begin{pmatrix}
			i \zeta_{2} \gamma_{-} \\
			- e^{- i \theta_{2}} \gamma_{+} \\
			e^{- i \theta_{2}} \\
			i \zeta_{2}
		\end{pmatrix};
		\hat{B}_{3} =
		\begin{pmatrix}
			i \\
			- \zeta_{1} e^{- i \theta_{1}} \\
			\zeta_{1} e^{- i \theta_{1}} \gamma_{-} \\
			i \gamma_{+}
		\end{pmatrix};
		\hat{B}_{4} =
		\begin{pmatrix}
			i \zeta_{2} e^{- i \theta_{2}} \gamma_{-} \\
			\gamma_{+} \\
			-1 \\
			i \zeta_{2} e^{- i \theta_{2}}
		\end{pmatrix}.
	\end{align}
	The scattering solutions satisfying the out-going boundary condition at $x=+\infty$ ($x=-\infty$), denoted by
	$\hat{\tilde{\psi}}_1(x')$ and $\hat{\tilde{\psi}}_2(x')$ ($\hat{\tilde{\psi}}_3(x')$ and $\hat{\tilde{\psi}}_4(x')$), are given in the $x'<0$ region as
	\begin{align}
		\hat{\tilde{\psi}}_{1}(x^{\prime}) &=
		\hat{B}_{1} e^{i k_{1} x^{\prime}} + \tilde{a}_{1} \hat{B}_{4} e^{i k_{2} x^{\prime}} + \tilde{b}_{1} \hat{B}_{3} e^{-i k_{1} x^{\prime}},
		\\
		\hat{\tilde{\psi}}_{2}(x^{\prime}) &=
		\hat{B}_{2} e^{-i k_{2} x^{\prime}} + \tilde{a}_{2} \hat{B}_{3} e^{-i k_{1} x^{\prime}} + \tilde{b}_{2} \hat{B}_{4} e^{i k_{2} x^{\prime}},
		\\
		\hat{\tilde{\psi}}_{3}(x^{\prime}) &=
		\tilde{c}_{3} \hat{B}_{3} e^{-i k_{1} x^{\prime}} + \tilde{d}_{3} \hat{B}_{4} e^{i k_{2} x^{\prime}},
		\\
		\hat{\tilde{\psi}}_{4}(x^{\prime}) &=
		\tilde{c}_{4} \hat{B}_{4} e^{i k_{2} x^{\prime}} + \tilde{d}_{4} \hat{B}_{3} e^{-i k_{1} x^{\prime}},
	\end{align}
	where $\tilde{c}_3$ ($\tilde{d}_4$) and $\tilde{d}_3$ ($\tilde{c}_4$) are the transmission coefficients for electron and hole, respectively,
	when electron (hole) is injected from the right.

	\subsection{Determining the coefficients}
	Green's function is constructed by making the linear combination of the outer products of the above obtained solutions,
	resulting in~\cite{PhysRev.175.559,PhilosTransRSocA.376.20150246}
	\begin{align}
		G^\mathrm{R}(x, x^{\prime}) =
		\begin{cases}
			\alpha_{1} \hat{\psi}_{1}(x) \hat{\tilde{\psi}}_{3}^{\mathsf{T}}(x^{\prime}) + \alpha_{2} \hat{\psi}_{1}(x) \hat{\tilde{\psi}}_{4}^{\mathsf{T}}(x^{\prime}) +
			\alpha_{3} \hat{\psi}_{2}(x) \hat{\tilde{\psi}}_{3}^{\mathsf{T}}(x^{\prime}) + \alpha_{4} \hat{\psi}_{2}(x) \hat{\tilde{\psi}}_{4}^{\mathsf{T}}(x^{\prime})
			& (x > x^{\prime}) \\
			\beta_{1}  \hat{\psi}_{3}(x) \hat{\tilde{\psi}}_{1}^{\mathsf{T}}(x^{\prime}) +  \beta_{2} \hat{\psi}_{4}(x) \hat{\tilde{\psi}}_{1}^{\mathsf{T}}(x^{\prime}) +
			\beta_{3}  \hat{\psi}_{3}(x) \hat{\tilde{\psi}}_{2}^{\mathsf{T}}(x^{\prime}) +  \beta_{4} \hat{\psi}_{4}(x) \hat{\tilde{\psi}}_{2}^{\mathsf{T}}(x^{\prime})
			& (x < x^{\prime})
		\end{cases}.
		\label{eq:retarded_green_function}
	\end{align}
	The boundary condition of Green's function at $x = x^{\prime}$ (Eq.~\eqref{eq:boundary_condition}) gives following set of equations
	\begin{align}
		\alpha_{1} \tilde{c}_{3} + \alpha_{2} \tilde{d}_{4} = \beta_{1} c_{3} + \beta_{2} d_{4} &=
		\frac{i}{2 v_{\mathrm{F}} \zeta_{1} \cos \theta_{1} (1 - \gamma_{+} \gamma_{-})}, \\
		\alpha_{3} \tilde{d}_{3} + \alpha_{4} \tilde{c}_{4} = \beta_{3} d_{3} + \beta_{4} c_{4} &=
		\frac{i}{2 v_{\mathrm{F}} \zeta_{2} \cos \theta_{2} (1 - \gamma_{+} \gamma_{-})}, \\
		\alpha_3 \tilde{c}_3 + \alpha_4 \tilde{d}_4 = \alpha_3 \tilde{c}_3 + \alpha_4 \tilde{d}_4 &=
		\beta_1 d_3 + \beta_2 c_4 = \beta_3 c_3 + \beta_4 d_4 = 0,
		\\
		(\alpha_{1} \tilde{c}_{3} + \alpha_{2} \tilde{d}_{4}) a_1 &= (\beta_{3} d_{3} + \beta_{4} c_{4}) \tilde{a}_2, \\
		(\alpha_{3} \tilde{d}_{3} + \alpha_{4} \tilde{c}_{4}) a_2 &= (\beta_{1} c_{3} + \beta_{2} d_{4}) \tilde{a}_1, \\
		(\alpha_{1} \tilde{c}_{3} + \alpha_{2} \tilde{d}_{4}) b_1 &= (\beta_{1} c_{3} + \beta_{2} d_{4}) \tilde{b}_1, \\
		(\alpha_{3} \tilde{d}_{3} + \alpha_{4} \tilde{c}_{4}) b_2 &= (\beta_{3} d_{3} + \beta_{4} c_{4}) \tilde{b}_2.
	\end{align}
	The first two equations give the definitions for $g_{1}$ and $g_{4}$ in the main text, respectively. Combining these equations, the Green's function is derived in the form of Eq.~\eqref{eq:green_function}.
\end{widetext}

\section{Josephson current}\label{sect:appendixB}
The calculation of Josephson current is done with evaluating the following quantities by using Green's function defined in the left superconducting region ($x<0$). The current satisfies the continuity equation
\begin{align}
	\frac{\partial}{\partial t} \rho + \frac{\partial}{\partial x} j + S = 0,
\end{align}
where $\rho$, $j$, and $S$ are the charge density, the current, and the source term operators, respectively.
They are defined as
\begin{align}
	\rho &= -e \left( c_{\uparrow}^{\dagger} c_{\uparrow} + c_{\downarrow}^{\dagger} c_{\downarrow}\right),
	\\
	j &= i e v_{\mathrm{F}} \left( c_{\downarrow}^{\dagger} c_{\uparrow} - c_{\uparrow}^{\dagger} c_{\downarrow} \right),
	\\
	S &= i e \Delta_{0} \left( c_{\downarrow} c_{\uparrow} - c_{\uparrow}^{\dagger} c_{\downarrow}^{\dagger}
	- c_{\uparrow} c_{\downarrow} + c_{\downarrow}^{\dagger} c_{\uparrow}^{\dagger} \right).
\end{align}
As we defined in the main text, $c_{s}^{\dagger}$ ($c_{s}$) is the Fourier transform of the creation (annihilation) operator $c_{\bm{k},s}$ of an electron with spin $s=\uparrow,\downarrow$ with respect to $x$.
The expectation values of them at $x$ are evaluated in terms of Green's function with the summation over Matsubara frequency and the wave number along the $y$ direction:
\begin{widetext}
	\begin{align}
		\langle j\rangle_{\omega_{n}>0} &= \frac{1}{\beta} \sum_{k_{y},\,\omega_n > 0} J_n \\
		J_n &= \frac{i e v_\mathrm{F}}{2} \left[ G_{12}(x, x, k_{y}, i \omega_n) - G_{21}(x, x, k_{y}, i \omega_n) + G_{34}(x, x, k_{y}, i \omega_n) - G_{43}(x, x, k_{y}, i \omega_n)\right] \\
		&= e \left( \gamma_+ e^{i (\theta_1 + \theta_2)} + \zeta_1 \zeta_2 \gamma_- \right)
		\left( \frac{i e^{-i \theta_1}}{2 \zeta_1 \cos \theta_1 (1 - \gamma_+ \gamma_-)} a_1
		- \frac{i e^{-i \theta_2}}{2 \zeta_2 \cos \theta_2 (1 - \gamma_+ \gamma_-)} a_2 \right)
		e^{i (k_2 - k_1) x}
		\\
		&= i e \frac{\gamma_+ e^{i (\theta_1 + \theta_2)} + \zeta_1 \zeta_2 \gamma_-}
		{2(1 - \gamma_+ \gamma_-)} \left( \frac{e^{-i \theta_1}}{\zeta_1 \cos \theta_1} a_1
		- \frac{e^{-i \theta_2}}{\zeta_2 \cos \theta_2} a_2 \right) e^{i(k_2 - k_1)x}
		\\
		&=
		\frac{e \Delta}{4\sqrt{\omega_n^{2} + \Delta_z^{2}}} \left\{ \left( 1 - \frac{m_z}{\mu} \right)
		e^{i(\theta_1 + \theta_2)} + \left( 1 + \frac{m_z}{\mu} \right) \zeta_1 \zeta_2 \right\}
		\left( \frac{e^{-i \theta_1}}{\zeta_1 \cos \theta_1} a_1 - \frac{e^{-i \theta_2}}{\zeta_2 \cos \theta_2} a_2 \right) e^{i(k_2 - k_1)x}
		\\
		&=
		\frac{e v_{\mathrm{F}} \Delta (k_1 + k_2) (e^{i \theta_1} \zeta_2 + e^{i \theta_2} \zeta_1)}
		{8 \mu \sqrt{\omega_n^{2} + \Delta_z^{2}}} \left( \frac{e^{-i \theta_1}}{\zeta_1 \cos \theta_1} a_1
		- \frac{e^{-i \theta_2}}{\zeta_2 \cos \theta_2} a_2 \right) e^{i(k_2 - k_1)x},
	\end{align}
	\begin{align*}
		\langle S\rangle_{\omega_{n}>0} &= \frac{1}{\beta} \sum_{k_{y},\,\omega_n > 0} S_n \\
		S_n &= i e \Delta \left[ G_{14}(x, x, k_{y}, i \omega_n) - G_{23}(x, x, k_{y}, i \omega_n) + G_{32}(x, x, k_{y}, i \omega_n) - G_{41}(x, x, k_{y}, i \omega_n) \right]
		\\
		&= ie \Delta \left[ a_1 \frac{i}{2 v_\mathrm{F} \zeta_1 \cos \theta_1} \left( e^{i(\theta_2 - \theta_1)} \zeta_1 + \zeta_2 \right)
		- a_2 \frac{i}{2 v_\mathrm{F} \zeta_2 \cos \theta_2} \left( e^{i(\theta_1 - \theta_2)} \zeta_2 + \zeta_1 \right) \right] e^{i (k_2 - k_1) x}
		\\
		&= - \frac{e \Delta}{2 v_\mathrm{F}} (e^{i \theta_1} \zeta_2 + e^{i \theta_2} \zeta_1)
		\left( \frac{e^{-i \theta_1}}{\zeta_1 \cos \theta_1} a_1 - \frac{e^{-i \theta_2}}{\zeta_2 \cos \theta_2} a_2 \right) e^{i(k_2 - k_1)x},
	\end{align*}
	and
	\begin{align}
		\int_{0}^{x} \langle S_n\rangle_{\omega_{n}>0} dx
		&= - \frac{e \Delta (e^{i \theta_1} \zeta_2 + e^{i \theta_2} \zeta_1)}
		{2 i v_\mathrm{F} (k_2 - k_1)} \left( \frac{e^{-i \theta_1}}{\zeta_1 \cos \theta_1} a_1
		- \frac{e^{-i \theta_2}}{\zeta_2 \cos \theta_2} a_2 \right) \left( e^{i(k_2 - k_1)x} - 1 \right)
		\\
		&=
		\frac{e v_{\mathrm{F}} \Delta (k_1 + k_2) (e^{i \theta_1} \zeta_2 + e^{i \theta_2} \zeta_1)}
		{8 \mu \sqrt{\omega_n^{2} + \Delta_z^{2}}} \left( \frac{e^{-i \theta_1}}{\zeta_1 \cos \theta_1} a_1
		- \frac{e^{-i \theta_2}}{\zeta_2 \cos \theta_2} a_2 \right)\left( 1 - e^{i(k_2 - k_1)x} \right),
	\end{align}
	obtaining
	\begin{align}
		I_{\omega_n > 0} &= \langle j\rangle_{\omega_{n}>0} + \int_{0}^{x} \langle S\rangle_{\omega_{n}>0} dx
		\\
		&= \frac{1}{\beta} \sum_{\omega_n > 0}
		\frac{e v_{\mathrm{F}} \Delta (k_1 + k_2) (e^{i \theta_1} \zeta_2 + e^{i \theta_2} \zeta_1)}{8 \mu \sqrt{\omega_n^{2} + \Delta_z^{2}}}
		\left( \frac{e^{-i \theta_1}}{\zeta_1 \cos \theta_1} a_1 - \frac{e^{-i \theta_2}}{\zeta_2 \cos \theta_2} a_2 \right).
	\end{align}
	The same calculation with the advanced Green's function should be performed for the summation $\omega_{n}<0$ and it turns out to be the same as the above equation. Therefore, the Josephson current is obtained as Eq.~\eqref{eq:josephson} in the main text.
	In the quasiclassical limit, the variables appearing in the eigenstates are approximated as
	\begin{align}
		Z_{1(2)\pm} &\simeq \mu \mp m_z \equiv Z_{\pm}, \\
		\zeta_{1(2)} &\simeq \sqrt{\frac{Z_+}{Z_-}} \equiv \zeta, \\
		k_{1(2)} &\simeq  k_\mathrm{F} \cos \theta \pm \frac{\mu \Omega_z}{v_\mathrm{F}^{2} k_\mathrm{F} \cos \theta},\quad k_\mathrm{F} \equiv \frac{\sqrt{\mu^{2} - m_z^{2}}}{v_\mathrm{F}}, \\
		\theta_1 &\simeq \theta_2 \simeq \theta,
	\end{align}
	resulting in Eq.~\eqref{eq:quasiclassical} in the main text.
\end{widetext}

%
\end{document}